\documentclass[amssymb,nobibnotes,aip,jcp,superscriptaddress,preprint]{revtex4}

\usepackage{graphicx}
\usepackage{dcolumn}
\usepackage{amsmath,amssymb,amsfonts}
\usepackage{epstopdf}
\usepackage{color}

\begin{document}
\title{Interaction between like-charged polyelectrolyte-colloid complexes in electrolyte solutions: a Monte Carlo simulation study in the Debye-H\"{u}ckel approximation}

\author{D. Truzzolillo}
\affiliation{Dipartimento di Fisica, Universita' di Roma "La Sapienza"\\
Piazzale A. Moro 5, I-00185 - Rome (Italy)  and F.O.R.T.H., Institute of Electronic Structure and Laser, Heraklion, Crete, Greece.}
\author{F. Bordi}
\affiliation{Dipartimento di Fisica and CNR-IPCF, Universita' di Roma "La Sapienza"\\
Piazzale A. Moro 5, I-00185 - Rome (Italy) }
\author{F. Sciortino}
\affiliation{Dipartimento di Fisica and CNR-ISC, Universita' di Roma "La Sapienza"\\
Piazzale A. Moro 5, I-00185 - Rome (Italy) }
\author{S. Sennato}
\affiliation{Dipartimento di Fisica, Universita' di Roma "La Sapienza"\\
Piazzale A. Moro 5, I-00185 - Rome (Italy) and CNISM}

\date{\today}

\begin{abstract} We study the effective interaction between differently charged polyelectrolyte-colloid complexes in electrolyte solutions via Monte Carlo simulations. These complexes are formed when short and flexible polyelectrolyte chains adsorb onto oppositely charged colloidal spheres, dispersed in an electrolyte solution. In our simulations the bending energy between adjacent monomers is small compared to the electrostatic energy, and the chains, once adsorbed, do not exchange with the solution, although they rearrange on the particles surface to accomodate further adsorbing chains or due to the electrostatic interaction with neighbor complexes. Rather unexpectedly,
when two interacting particles approach each others, the rearrangement of the surface charge distribution invariably produces anti-parallel dipolar doublets, that invert their orientation at the isoelectric point. These findings clearly rule out a contribution of dipole-dipole interactions to the observed attractive interaction between the complexes, pointing out that such suspensions can not be considered dipolar fluids. On varying the ionic strength of the electrolyte, we find that a screening length $\kappa^{-1}$, short compared with the size of the colloidal particles, is required in order to observe the attraction between like charged complexes due to the non-uniform distribution of the electric charge on their surface ('patch attraction').
On the other hand, by changing the polyelectrolyte/particle charge ratio, $\xi_s$, the interaction between like-charged polyelectrolyte-decorated (pd) particles, at short separations,  evolves from purely repulsive to strongly attractive.
Hence, the effective interaction between the complexes is characterized by a potential barrier, whose height depends on the net charge and on the non-uniformity of their surface charge distribution.\end{abstract}
\maketitle

\section{Introduction}
 Understanding complexation of linear polyelectrolyte chains with oppositely charged particles
  has  high potential for applications in nanoscience and bio-technology \cite{Bordi09,Yaroslavov07,Yaroslavov07b,Volodkin06,Safinya01,Radler97}. In a number of technological processes and biological systems, linear polyelectrolytes associate with a range of different nanoscopic particles or supramolecular assemblies, including inorganic particles, protein assemblies, micelles and liposomes, forming in aqueous media charged colloidal suspensions. DNA, which is a polyelectrolyte, provides a typical example of the relevance of these interactions. Within the cell nucleus the very long DNA chains are packed and ordered into much smaller structural units (nucleosomes) due to their interaction with strongly alkaline proteins, the histones, acting as 'spools' around which DNA winds.\\
More in general, the long-range character of the electrostatic interaction, and its interplay, in a polyelectrolyte solution, with entropic and conformational effects, result in a very complex and interesting phenomenology \cite{Dobrynin05,Likos04}. Particularly, when polyelectrolytes adsorb on an oppositely charged extended surface, entropic, geometric and correlation effects play a pivotal role.\\
There is now accumulating evidence that the polyelectrolyte adsorption, due to the repulsion between the like-charged chains, occurs in a correlated manner \cite{Dobrynin00}, and that the intriguing phenomenon of the overcharging \cite{Henon04,Dobrynin01,Grosberg02,Lenz08}, when more polyelectrolyte than would be needed to neutralize the surface adsorbs, so that the sign of the net charge of the 'decorated' surface is inverted, can be understood in terms of correlated adsorption \cite{Grosberg02} and 'charge fractionalization' \cite{Nguyen02c,Nguyen02d}. The basic idea of 'fractionalization' is that by forming loops or leaving dangling  ends, the adsorbed chains gain some conformational entropy. The vacancies left by these defects can locally be large enough to drive closer to the surface oncoming polyelectrolytes, that may adsorb, again with some loop and dangling ends (above the surface). As a result, instead of having a Z-charged chain in solution and the surface neutralized by a uniform coating of ordered chains laying flat on it, Z 'disconnected' charges appear 'protruding' from the adsorbed polyelectrolyte layer: the charge of the polyelectrolyte is 'fractionalized' along the surface. This is an energetically favored configuration for the system, that gains some conformational entropy of the adsorbed chains and also, by 'diluting' its charge on the whole surface, part of the electrostatic self-energy of the Z-charged polyelectrolyte.\\
By increasing the polyelectrolyte/particle charge ratio, the net charge of the decorated particles changes progressively from the value of the \emph{bare} particle to the maximum 'overcharging' inverting its sign at the isoelectric point.\\
The correlated adsorption of the chains also results in a non-uniform distribution of the electrostatic charge on the surface of the \emph{decorated} particles. As a consequence of this non-uniformity, an attractive interaction can be observed between like-charged particles \cite{Velegol01,Velegol02,Mukherjee04}.\\
This attractive component of the inter-particle potential, and the progressive variation of the net charge of the decorated particles, down to the neutralization point and up again to the maximum overcharging, that modulate the electrostatic repulsion, combine together yielding a peculiar phenomenon of 're-entrant condensation' of the polyelectrolyte-decorated particles. As the polyelectrolyte/particle ratio is increased, associated to the progressive reduction of the net charge of the primary polyelectrolyte-decorated particles (pd-particles), larger and larger clusters are observed. Close to the isoelectric condition the aggregates reach their maximum size, while beyond this point any further increase of the polyelectrolyte-particle charge ratio causes the formation of aggregates whose size is progressively reduced.\\
Eventually, when the surface of the particles is completely saturated by the adsorbed polyelectrolyte, i.e. the overcharging has reached its maximum extent, the size of the particles in the suspension equals again the size of the original colloidal particles, plus a thin layer of adsorbed polymer\cite{Volodkin06,Bordi04,Yaroslavov07b}. From here on, by further increasing the polymer/particle ratio, the excess polyelectrolyte, that does not adsorb any more on the particles, remains 'freely' dissolved, contributing to the overall ionic strength of the solution.\\

This intriguing phenomenology has been observed in a variety of polyelectrolyte-colloid systems dispersed in aqueous solutions such as polyelectrolyte-micelle complexes,\cite{Wang00} latex particles \cite{Gillies07, Keren02}, dendrimers \cite{Kabanov00}, ferric oxide particles \cite{Radeva08}, phospholipid vesicles (liposomes) \cite{Bordi06, Volodkin07, Radler98}, 'hybrid niosome' vesicles \cite{Sennato08b}.\\

Velegol and Thwar \cite{Velegol01} have recently suggested an analytical model (based on the Derjaguin approximation and on an extension of the Hogg-Healy-Fuerstenau model \cite{Hogg66}) for the potential of mean force (VT-potential hereafter) between non-uniformly charged surfaces (planar and spherical), showing that the interaction between non-uniformly charged surfaces results in an inter-particle potential that, even in the case of like-charged particles, has an attractive component.\\
Accurate simulations on a system composed by spherical particles interacting via VT-potential \cite{Truzzolillo08} showed that the overall phenomenology is, at least qualitatively, well-described by this model. In particular, the balance of attractive and repulsive electrostatic interactions yields a potential barrier whose height increases with the size of the aggregates, justifying the stabilization of the aggregates. Aggregates  stop growing after reaching a characteristic size, determined by the net charge and the degree of non-uniformity. Such behavior is consistent with the observed 'reentrant condensation' and with the thermally activated character of the aggregation~\cite{Truzzolillo08,Sennato08}.\\

Most of the previous works on polyelectrolyte-colloid complexes focused on the complexation behavior, i.e on the adsorption of PE chains on a single oppositely charged colloidal particle \cite{Granfeldt91, Wallin96,Wallin98, Jonsson01, Jonsson01b, Nguyen01c, Netz99} or viceversa of one long PE on many particles \cite{Jonsson01, Jonsson01b}, being mainly concerned with the overcharging phenomenon.
A few authors discussed the interactions between two macroions in the presence of one long (compared to the particle size) polyelectrolyte chain \cite{Dzubiella03, Granfeldt91} or several short chains \cite{Nguyen01a,Podgornik95} in different environments.
Here we mainly focus instead on the effective interaction between decorated particles, considering the complexes formed by one spherical macroion and several PE-chains whose length is small compared with the particle size. We investigate the effective interaction between two of such decorated colloids in and electrolyte at varying the screening length, $\kappa^{-1}$, and the polyelectrolyte concentration. In our simulations we reproduce the conditions that are typical for the polyelectrolyte-liposome systems employed in several experimental works \cite{Bordi04,Bordi04b,Bordi04c,Bordi05,Bordi05b,Bordi06,Sennato08b} and that show an interesting potential for biotechnological applications as drug carriers for intra-cellular drug delivery \cite{Bordi09}. To this aim we perform Monte Carlo (MC) simulations of PE-decorated particles, for different polyelectrolyte/particle charge ratios, in the limit of  bending energy small compared to the electrostatic interaction (rather flexible chains).

Dzubiella et al. \cite{Dzubiella03} showed that, for rather small colloidal particles (with a size not exceeding ten times the radius of a polyelectrolyte monomer) and almost neutral complexes,
the shape of the interaction potential as a function of the inter-particle distance is only minimally affected by the ionic strength of the solution. On the contrary, here we show that, for short chains and small monomers and for non-neutral complexes, this parameter plays, through the screening length $\kappa$, a key role, modulating the balance of repulsive (due to the net charge) and attractive (patch) interactions. We find that
the addition of the polyelectrolyte chains on the macroions' surfaces switches the short distance interaction from repulsive to attractive, yielding a potential barrier modulated by the net charge and by the non-uniformity of its distribution. The nature of the short range attractive interaction between the decorated particles will be further clarified by calculating their dipole moments. Rather unexpectedly, the dipole moments of approaching particles remain anti-parallel, ruling out a contribution of dipolar interactions to the attractive component of the inter-particle forces, and showing that the condensation of the complexes is only controlled by the charge mismatch between the two opposing surfaces.

\section{The model}\label{model}
We perform MC-simulations adopting the scheme proposed by Kremer et al. \cite{Grest87, Grest94} for polyelectrolyte chains, and the  pearl necklace model to generate off-lattice 3-dimensional polymer chains represented as a succession of $N=20$ freely jointed Lennard-Jones spheres (beads). Each bead is a physical monomer with one positive elementary charge positioned at its center, and a diameter equal to one Bjerrum length, $l_B= e^2/\epsilon k_B T$, the distance between two point charges $e$ where their electrostatic interaction energy in a medium with dielectric permittivity $\epsilon$ reduces to the thermal energy $k_B T$ ($7.14$ {\AA} in water at 25 $^\circ$C).\\
The fraction of ionized monomers $f$ is set to $1$ whereas the bond length is a fluctuating quantity whose oscillation is tuned by the binding potential. Assuming good solvent conditions, a shifted Lennard-Jones (LJ) potential is used to describe the purely repulsive excluded volume interaction between the $N$ monomers:

\begin{equation}\label{LJ} V_{LJ}(r) =
\left\{
\begin{array}{ll}
 4\epsilon_{LJ}\left[\left(\frac{2R_{m}}{r}\right)^{12}-\left(\frac{2R_{m}}{r}\right)^6
 +\frac{1}{4}\right] & r\leq 2^{\frac{5}{6}}R_{m} \\
 0 & r\geq 2^{\frac{5}{6}}R_{m}
\end{array}
\right.
\end{equation}

where $R_m$ is the radius of the beads (the microscopic length scale), $r$ their relative distance,  and $\epsilon_{LJ}$ sets the energy scale. The connectivity of the bonded monomers is assured by a finite extension nonlinear elastic (FENE) potential \cite{Dzubiella03} acting between neighboring beads
\begin{equation} V_{LJ}(r) =
\left\{
\begin{array}{ll}
 -\frac{1}{2}k_F\left(\frac{r_{max}}{2R_m}\right)^2\ln\left[1- \left(\frac{r}{r_{max}}\right)^2\right]
 & r<r_{max}\\
\infty & r>r_{max}
\end{array}
\right.
\end{equation}

where $k_F$ denotes the spring constant. Following Dzubiella et al. \cite{Dzubiella03} we have chosen $k_F=7.0\epsilon_{LJ}$, the maximal relative displacement of two neighboring beads  $r_{max}=4R_m$ and the L-J parameter  $\epsilon_{LJ}=1.2 k_BT$ \cite{Grest86,Jusufi02b,Dzubiella03}. These values of the parameters prevent the chains from crossing \cite{Grest86}.\\

Finally, electrostatic interactions between monomers have been taken into account via the Debye-H\"{u}ckel potential
\begin{equation}\label{DH}
 V_{DH}(r)=\frac{z^2e^2 \exp\left[-\kappa r\right]}{4\pi\epsilon r}
\end{equation}

where $z$=$1$ is the monomer valence, $e$ the elementary charge and $\epsilon$ is the dielectric permittivity of the medium, set to $7.08 \cdot 10^{-10}\frac{C^2}{Jm}$. $\kappa^{-1}$ is the Debye screening length, related to the electrolyte concentration as follows

\begin{equation}\label{kappaquadro}
 \kappa^2=1000 \cdot e^2 N_A\sum_i \frac{z_i^2 C_i}{\epsilon k_B T}
\end{equation}

where where $N_A$ is the Avogadro's number, $z_i$ the valence of the ionic species, $C_i$ the electrolyte concentration in mol/l, and $k_B T$ the thermal energy. The solvent is only taken into account via the dielectric constant. It must be noted that in the absence of any added salt, the system contains the small counterions resulting from the ionization of both the chains and the colloidal particles which, as stated above, is assumed to be complete. However, also these small counterions are not considered explicitly in the simulations but only taken into account via the screening constant $\kappa^{-1}$.

We do not add  a bending term to the total pair interaction energy between adjacent monomers.

Using these parameters, the bond-length $b$, defined as the distance between the centers of two consecutive monomers, has a weak dependence on ionic strength ($b$ varies from $0.84$ nm to $0.82$ nm on passing from $\kappa=1$ nm$^{-1}$ to  $\kappa=0.1$ nm$^{-1}$ ). In this conditions, the value of the Manning condensation parameter $l_B/b$ \cite{Manning69,Manning78} is less than $1$, justifying the choice $f = 1$ (complete ionization). Hence, the total valence of one PE is simply given by $Z_p= N z$.

The interactions between each monomer and spherical macroions are modeled via Debye-Huckel and truncated LJ potentials as follows:

\begin{equation}\label{DH1}
 V_{DH}(r)=\frac{Zze^2\exp\left[-\kappa(r-R_c)\right]}{4\pi\epsilon r (1+\kappa R_c)}
\end{equation}

\begin{equation}\label{LJ1} V_{LJ}(r) =
\left\{
\begin{array}{ll}
 4\tilde{\epsilon}_{LJ}\left[\left(\frac{R_{m}+R_c}{r}\right)^{12}-\left(\frac{R_{m}+R_c}{r}\right)^6 +\frac{1}{4}\right] & r\leq 2^{\frac{1}{6}}(R_{m}+R_c) \\
 0 & r\geq 2^{\frac{1}{6}}(R_{m}+R_c)
\end{array}
\right.
\end{equation}

where $R_c$ is the radius of spherical macroions. The first potential (Eq. \ref{DH1}) drives the adsorption process and determines the internal structure of the PE-layer, the second one (Eq. \ref{LJ1}), where $\tilde{\epsilon}_{LJ}$ is set to $10\epsilon_{LJ}$, reproduces the strong excluded volume effects due to the steric interaction between the monomers and the macroion.

In our simulations we used, for different purposes, spherical particles with radii $R_c=3.41$ nm and $R_c=5$ nm, as will be shown in the next section.\\

In an attempt to reproduce what  occurs experimentally when some oppositely charged polyelectrolyte is added to a colloidal aqueous suspension~\cite{Bordi09}, we started simulating the formation of a decorated particle. A colloidal particle
was located at the center of a cubic box of side $L_{box}=60$ nm. Then $N_c$ chains were randomly  placed within the box, outside the excluded volume of the central sphere.
The macroion was kept fixed, while the polymers where moved according to
 the standard MC Metropolis algorithm\cite{Allen87}. Simulations have been carried out for a long equilibration time, typically of the order of $10^5$ MC steps, where one MC step consists of a trial translational move of every monomer. Periodic boundary condition have been applied.  In the equilibrium final configuration, $N_c$
 chains are adsorbed on the oppositely charged spheres.
 The  charge ratio $\xi_s=|fNN_c/Z|$, where  $Z$ the valence of the spherical macroion,
 provides a measure of the  net charge of the complex.
This preparation procedure has been carried out for each set of parameters used in the simulations.

Once a single complex, a pd-particle, was formed, we duplicated the whole structure (bare colloid + adsorbed PE chains) simply by translation. Then, to calculate the interaction energy \cite{Dzubiella03,Dzubiella00}, the system was equilibrated, keeping fixed the macroions, at various distances. The surface-to-surface distance between the
original particle and the copy was varied from $50$ nm to $0$ nm in the first series of simulations (for the longer screen distance, $\kappa^{-1}$= $10$ nm), and from $10$ nm to $0$ nm
in a second series ($\kappa^{-1}\leq2$ nm).\\
The electrostatic interaction between the macroions  was taken into account using the Debye-H\"{u}ckel potential \cite{Dzubiella03}

\begin{equation}\label{DHbare}
 V_{DH}(r)=\frac{Zze^2\exp\left[-\kappa(r-2 R_c)\right]}{4\pi\epsilon r (1+\kappa R_c)^2}
\end{equation}

while the excluded volume interaction was again modeled via a LJ potential, as in Eq. \ref{LJ1}, where $R_m$ has to be substituted by $R_c$.
All the simulations involving two pd-particles (sections \ref{thickness} and \ref{salt}) were performed using a cubic box much larger than the typical 'interaction volume' defined by the couple of particles; thus we did not need to apply periodic boundary conditions.\\
To favor the rearrangement of the chains toward the global energy minimum  (at fixed
macroion-macroion distance) we adopted two kinds of trial moves:
\begin{itemize}
  \item translational moves of single monomers,
  \item orientational moves of the whole polyelectrolyte layer around the macroion center.
\end{itemize}

The diffusion process along the axis connecting the macroions' centers depends on the net charge of each complex and on reciprocal distance, as we shall see studying the polarization effects.
One orientational move is performed randomly and, on average, every $5$ translational MC-steps. When such move is performed, each spherical PE-layer is rotated around the center of the respective adsorbing macroion.
Both the translational and orientational moves were applied with an acceptance ratio of $50$ $\%$.

The two-step procedure (formation of a decorated particle plus duplication) employed to obtain a couple of interacting complexes, can be justified considering that the diffusion of the colloidal particles and the adsorption process occur on two different time scales. The huge surface/volume ratio which  characterize colloids, and the distribution of this surface within the whole volume of the host phase, dramatically  speeds the adsorption process. Upon mixing the polyelectrolyte solution and the particle suspension, the time requested to the polyelectrolyte chains to reach by diffusion the particles' adsorbing surface is significantly smaller than the diffusion time of the macroions and hence the effective interaction is essentially between already fully decorated particles.
This conjecture, which appears in itself reasonable, has been recently substantiated with experimental evidence by Volodkin et al.\cite{Volodkin06}. By using different mixing protocols, varying the agitation speed and the order of mixing, these authors showed that the polyelectrolyte adsorption is almost immediate when compared with the characteristic times of the aggregation process, that are typical of the particles' diffusion.\\

\section{Polyelectrolyte adsorption: the effects of salt and charge ratio}\label{thickness}

The adsorbed polyelectrolyte (PE) chains play in principle a double role in determining the overall mean-force potential between the decorated particles: by building up a non zero dipole moment they can contribute repulsive or attractive components, moreover, at a sufficient short distance they produce a soft repulsive interaction. Both these effects depend on the distribution and on the conformation of the adsorbed chains. Thus, to gain insight on how the PE-layer of a non-neutral complex reacts to the presence of an approaching pd-particle, for different screening lengths $\kappa^{-1}$, is particularly important.
The Debye-Huckel potential, accounting for the screening created by the presence of small ions in the solution, greatly simplifies the study of the effect of changes of the ionic strength on the PE-layer thickness.\\

The typical value for  the macroion charge density in cationic liposomes is $\Sigma_{D}=1.67$ nm$^{-2}$ \cite{Bordi03}. We have selected this value for our study, resulting in a  macroion charge of $Z=-244$ for $R_c=3.41$ nm and $Z=-523$ for $R_c=5$ nm.
 In particular, the effect on the PE-layer of the screening length is here investigated
  using macroions with $R_c=3.41$ nm, and a surface charge ratio equal to $0.57$, equivalent to $7$ chains on each macroion (see the snapshot in the upper panel of Figure \ref{kappa1}).
\begin{figure}[htbp]
\begin{center}
  \includegraphics[width=8.5cm]{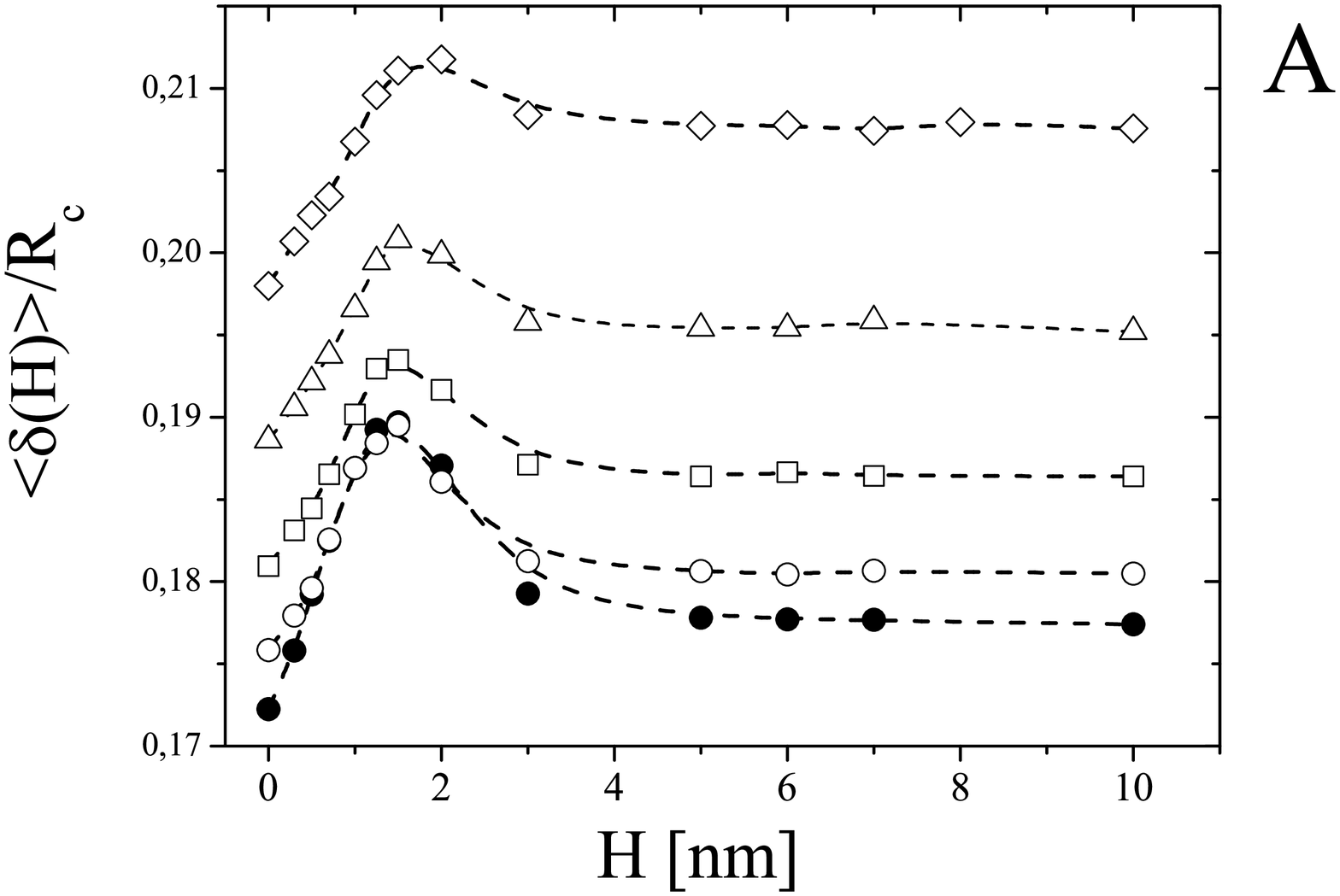}
  \includegraphics[width=8.5cm]{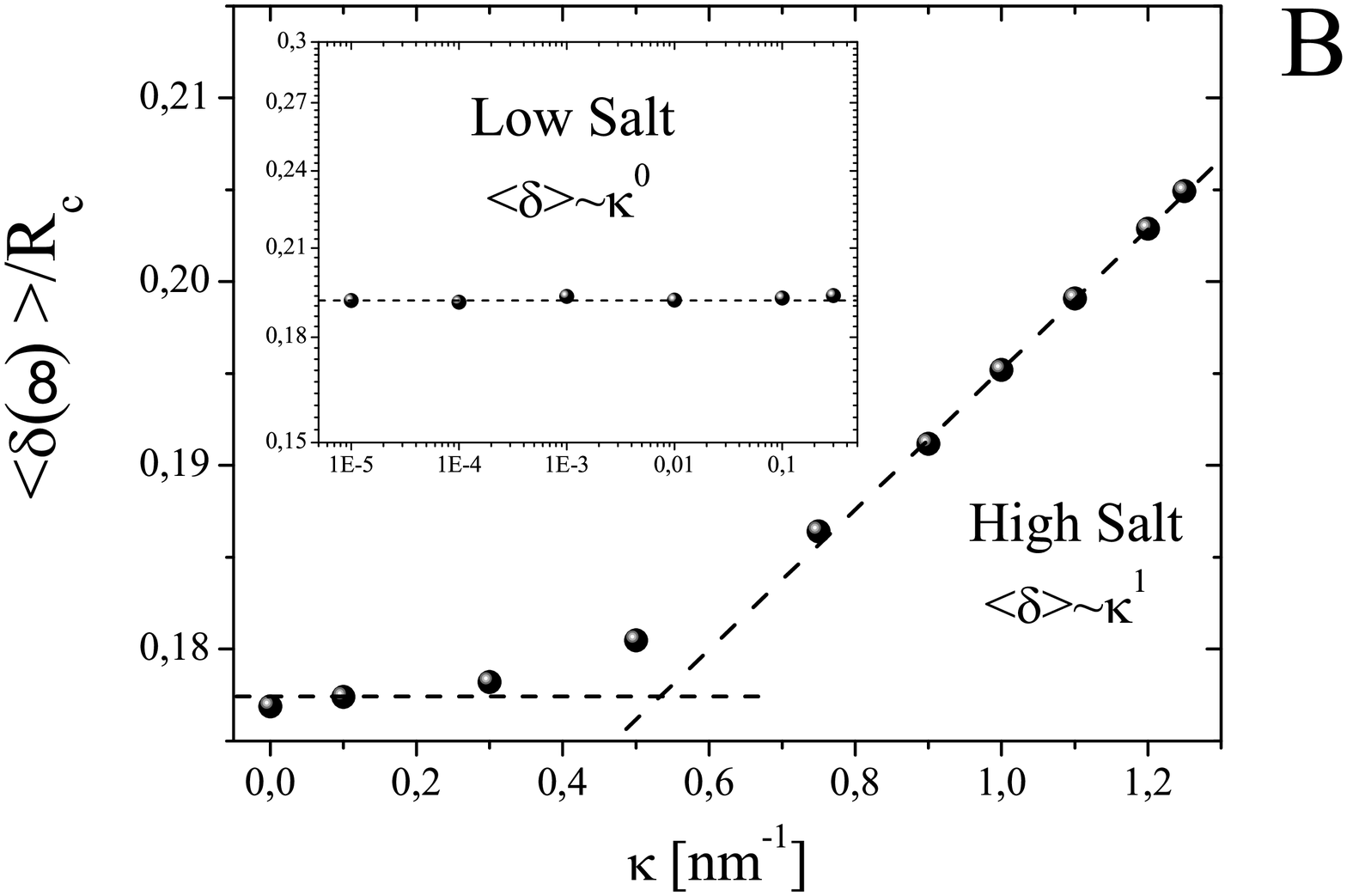}\\
  \caption{\textbf{PANEL A}: Averaged mean distance between the monomers and the bare macroion surface as a function of the distance $H$ between two approaching macroions. Simulations are carried out for PE-colloid complexes with $R_c=3.41$ nm, $\xi_s=0.57$ and different values of $\kappa$. Black filled points $\rightarrow$ $\kappa=0.1$ nm$^{-1}$;  circles $\rightarrow$ $\kappa=0.5$ nm$^{-1}$; squares $\rightarrow$ $\kappa=0.75$ nm$^{-1}$; triangles $\rightarrow$ $\kappa=1$ nm$^{-1}$; lozenges $\rightarrow$ $\kappa=1.25$ nm$^{-1}$. \textbf{PANEL B}: Averaged mean distance between the monomers and the bare macroion surface as a function of $\kappa$ for isolated complexes. \textbf{Inset}: Average mean distance $\langle\delta\rangle$ for $0$ nm$^{-1}$$<\kappa<0.3$ nm$^{-1}$ in a log-log scale.
  }\label{dist_coll_k}
  \end{center}
\end{figure}

We calculated the thickness $\langle\delta\rangle$ of the PE layer as the mean distance between the monomers and the macroion surface. For different values of $\kappa$, a common behavior of the thickness $\langle\delta\rangle$ as function of the distance $H$ between two complexes is recovered.
Figure \ref{dist_coll_k} (panel A) shows that the  approach of two complexes induces a weak but definite deformation of the polyelectrolyte adsorbed layer. When the distance between the two complexes becomes smaller than $\approx 4 \, nm$, the adsorbed monomers begin to be attracted by the field generated by the approaching complex. The resulting increase of their mean distance from the macroion surface, the layer 'swelling', depends on the inverse screening length $\kappa$. But, as expected, the distance where this increase appears, does not depend on $\kappa$ (both the monomer-colloid adsorption energy and the attraction induced by the approaching complex scale with the same $\kappa$).
At much shorter distances ($H<1$ nm) the layers on the approaching complexes become thinner, each layer being squeezed by its homologous on the other complex.
The position of the maximum of $\langle \delta(H)\rangle$ can be considered the effective range of the repulsive steric interaction between the complexes that, as will be shown in Fig. \ref{kappa1}, is $\kappa$-independent, at least in the range of Debye constant values investigated.\\

For comparison, Fig.~\ref{dist_coll_k} (panel B) shows the variation of the adsorbed layer thickness $\langle\delta\rangle$ of isolated pd-particles. The dependence of $\langle\delta\rangle$ on $\kappa$  points out that for the chosen values  of the surface charge density on the macroions, $\Sigma$, and of the valence of monomers $z$, the system is in the so called 'screening-reduced' regime, as predicted by previous analytical and numerical studies for electrostatically driven polymer adsorption \cite{Ciferri05,Steeg92}. Our data are consistent with the scaling laws obtained by Borukhov et al. \cite{Borukhov98} via mean field arguments for the dependence of the PE-layer thickness on salt concentration: $\langle\delta\rangle \sim \kappa$ for high values of $\kappa$ below the desorption threshold ($\kappa_{de}\simeq 1.5 nm^{-1}$), while for low salt concentration the thickness tends to be unperturbed by the increase of the screening, i.e., $\langle\delta\rangle \sim \kappa^0$ (see the inset of figure \ref{dist_coll_k} - panel B).\\

We also investigated the effect on the PE-layer thickness of the polyelectrolyte-particle charge ratio.
\begin{figure}[htbp]
\begin{center}
  \includegraphics[width=8cm]{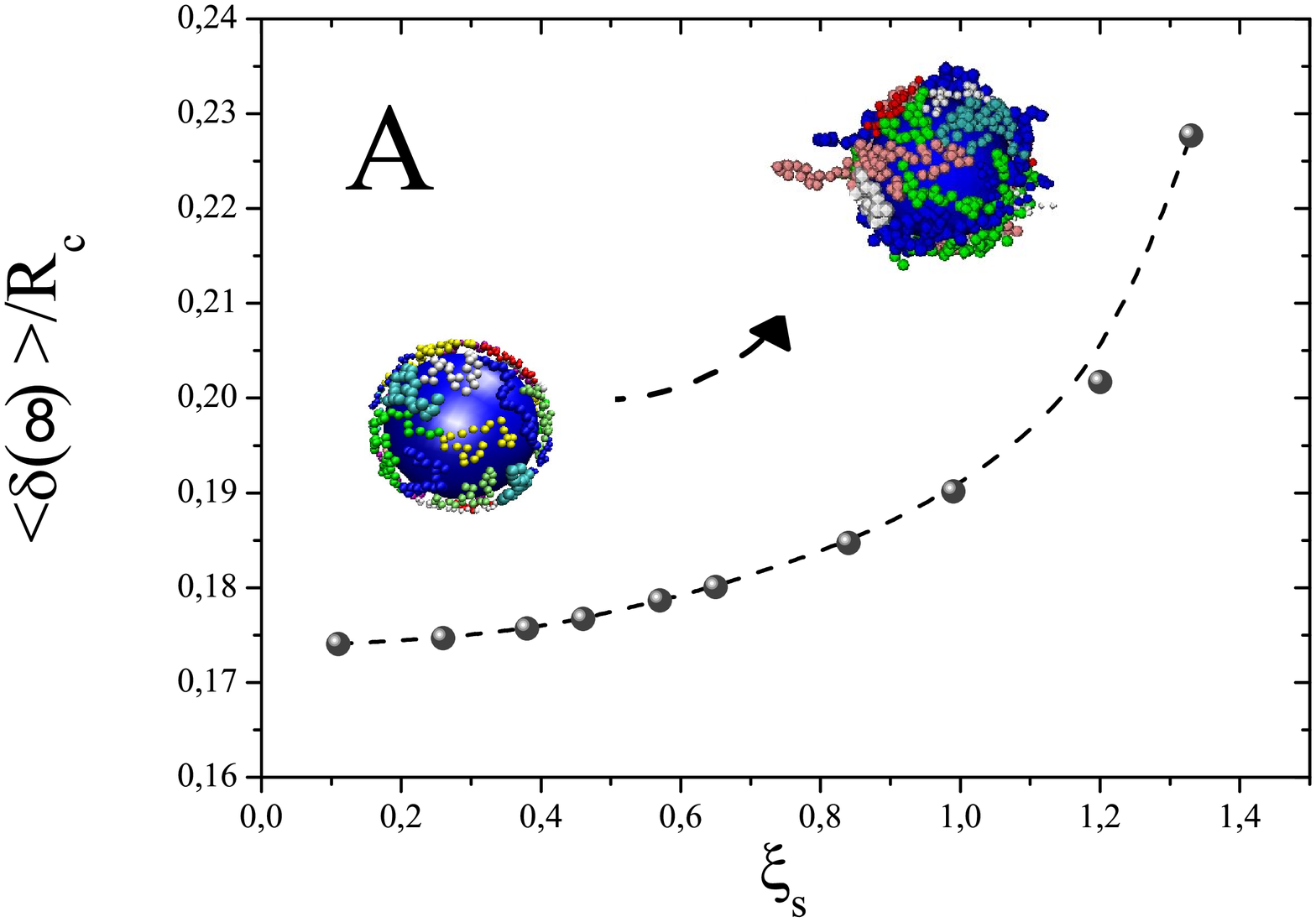}
  \includegraphics[width=8cm]{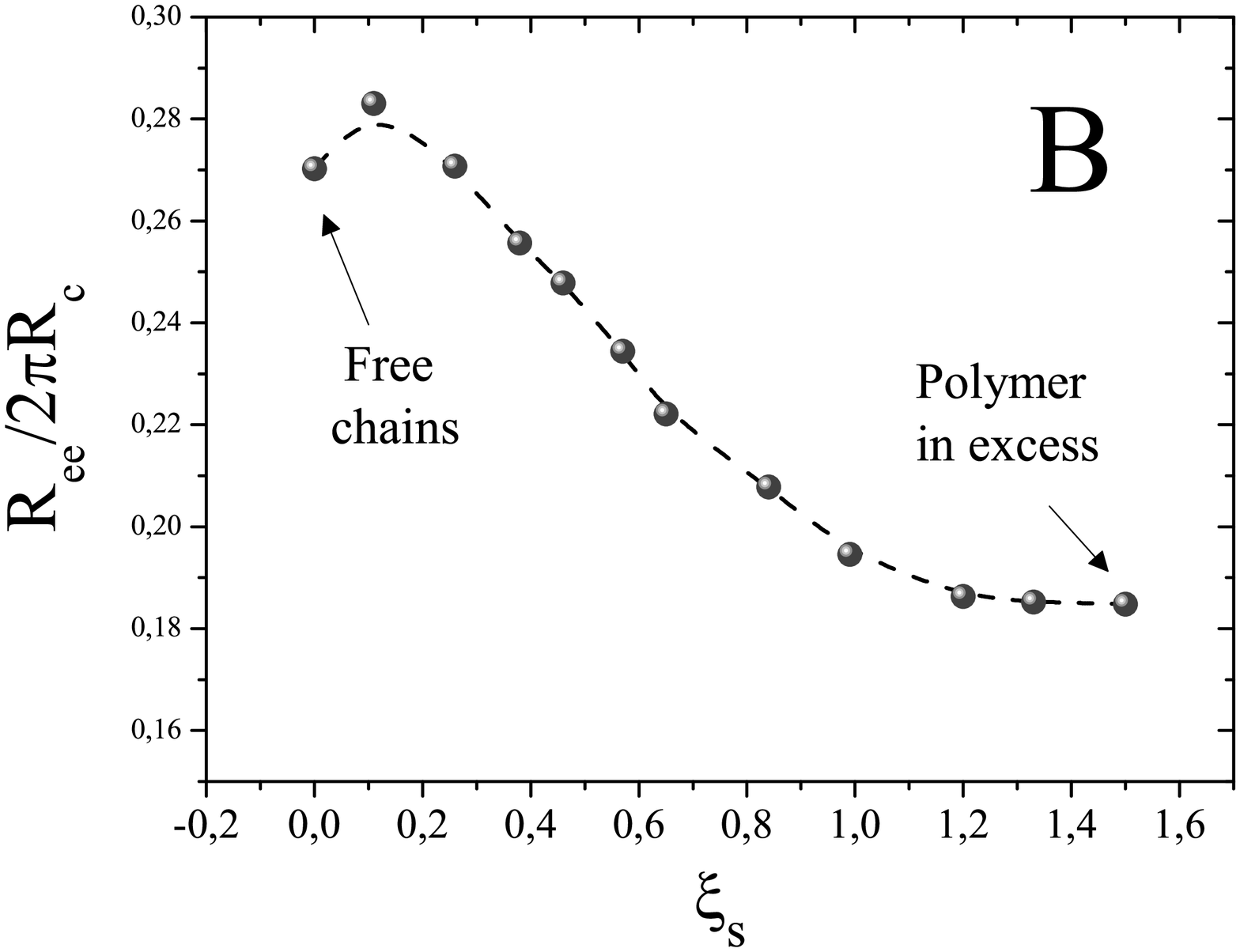}\\
  \caption{Averaged monomer-colloid distance (panel A) and average end-to-end distance of chains (panel B) for different charge ratio $\xi_s$, $\kappa=$ $1$ nm $^{-1}$, $R_c=5$ nm$^{-1}$. Dashed lines are guide to the eye only.
  }\label{chains_xi}
  \end{center}
\end{figure}
In a series of simulations we fixed the radius of the macroions to $R_c=5$ nm and the Debye constant to $\kappa=1$ nm$^{-1}$,  and added a different number of PE-chains so that the polyelectrolyte/macroion charge ratio varies from $\xi_s=0.11$ ($3$ chains) to $\xi_s=1.5$ ($39$ chains), where the polymer is in large excess.\\
The increased roughness of the PE-layer and the consequent increment of its thickness (see Fig. \ref{chains_xi}, panel A) for relatively high values of $\xi_s$ is a direct consequence of the lateral repulsion between the chains. The strong non linear behavior of $\langle\delta\rangle$ is due to the presence of a disordered three-dimensional arrangement of the chains, that do not lay flat on the surface but adopt a conformation characterized by loops and dangling ends.\\
To estimate a typical size of the adsorbed coils we calculate the averaged end-to-end distance
\begin{equation}\label{Ree}
R_{ee}=\left[ \langle \frac{1}{N_c}\sum_{i=1}^{N_c}(r_i(0)-r_i(N))^2\rangle\right]^{\frac{1}{2}}
\end{equation}
where $r_i(0)$ and $r_i(N)$ are the positions of the first and last monomer along the i$_{th}$ chain, respectively.\\
Figure \ref{chains_xi} (panel B) shows that at very low polyelectrolyte content, due to the steric constraint contributed by the spherical macroion, and to the negligible reciprocal interaction, the chains assume a rather extended conformation. The addition of more chains, due to the increased inter-chain repulsions, results in a reduction of the size of the adsorbed coils. This effect is, of course, directly correlated to the stiffness of chains: the introduction of an elastic bending energy would reduce the effect of lateral interactions on the coil size, that becomes negligible in the limit of rigid polymers. At higher polyelectrolyte content the shrinking of the chains levels off and large loops and dangling tails protruding from the surface appear, after that the condition of excess polyelectrolyte is reached (chains do not adsorb any more). The passage from a 'flat adsorption' to a 3D one, results in an extension of the steric repulsion range, that determines the shift of the global minimum of the effective interaction towards higher values of the inter-particle distance $H$.

\section{Effective interaction}\label{salt}
For each value of $\kappa$ and $\xi_s$, we calculated the mean force potential between two complexes at a distance $H$ by averaging the variation of the total energy $E$ of the system when the distance between the particles is reduced from infinity to $H$
\begin{equation}\label{pairpot1}
     \Phi(H)=\langle E(H)-E(\infty)\rangle
\end{equation}
\begin{figure}[htbp]
\begin{center}
  \includegraphics[width=8.5cm]{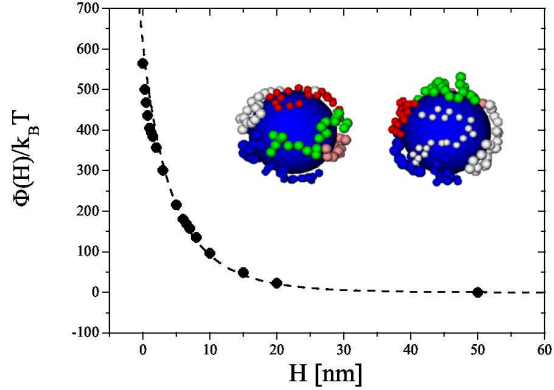}\\
  \caption{Mean force potential between two spherical PE-colloid complexes with $R_c=3.41$ nm $\xi_s=0.57$ and $\kappa=1$ nm $^{-1}$. The dashed line represents the potential calculated from Eq. \ref{DH2}. The snapshot shows two interacting pd-particles for $\xi_s=0.57$, $H=5$ nm and $R_c=3.41$ nm}\label{kappa1}
  \end{center}
\end{figure}
As a first step, we investigated the role of the screening on the interaction between non neutral complexes (the parameters are the same as in section \ref{thickness}, with $\xi_s=0.57$, $R_c=3.41$ nm and $7$ chains adsorbed on each macroion).\\
For large enough screening lengths, i.e. when $\kappa R_c<1$, the decorated particles are able to 'see' each other completely, in other words all the charges on the two complexes interact with each other, and all of them contribute to the effective pair potential. In these conditions the interaction between the pd-particles could be described as a \emph{"particle-particle"} interaction. Figure \ref{kappa1} shows the behavior of the mean force potential as a function of the inter-particle distance in these conditions, i.e. for $\kappa=0.1$ nm$^{-1}$ ($\kappa R_c=0.34$): the interaction is purely repulsive. The short range attractive component due to the matching between oppositely charge patches on the approaching particles does not balance the strong repulsion between the like-charged complexes.\\
In fact, the spatial profile of the mean force potential obtained can be exactly superimposed on the Debye-Huckel potential
\begin{equation}\label{DH2}
 V_{DH}(H)=\frac{(Z-\xi_s Z)^2e^2 \exp\left[-\kappa H\right]}{4\pi\epsilon (H+2R_c)(1+\kappa R_c)^2}
\end{equation}
where $Z-\xi_s Z$ is the net charge of the complexes (Figure \ref{kappa1}). As expected, for low screening conditions, i.e. $\kappa(R_c+\delta)<1$, where $\delta$ is the PE-layer thickness and  $\delta/R_c \ll 1$, the interaction is totally governed by the monopole-monopole repulsion. In this regime the non-uniformity of the charge distribution on the particles' surface has negligible effects and the interaction can be simply described as Debye-Huckel.\\

\begin{figure}[htbp]
\begin{center}
  \includegraphics[width=9cm]{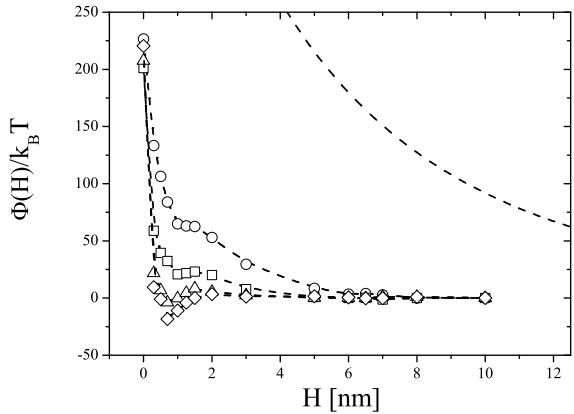}\\
  \caption{Total mean force potential between two spherical pd-particles with $R_c=3.41$ nm and $\xi_s=0.57$ for different values of $\kappa$. (o) $\kappa=0.5$ nm$^{-1}$; ($\Box$) $\kappa=0.75$ nm$^{-1}$; ($\triangle$) $\kappa=1$ nm$^{-1}$; ($\diamondsuit$) $\kappa=1.25$ nm$^{-1}$.
  The dashed line corresponds to the Debye-Huckel potential (Eq. \ref{DH2}) shown in Fig. \ref{kappa1} for $\kappa=0.1$ nm$^{-1}$.
  }\label{kappa2}
  \end{center}
\end{figure}

On the contrary, in high screening regime, due to the charge non-uniformity, a short range electrostatic attraction appears, which can not be justified within the simple Poisson Boltzmann theory for like charged colloids.\\
We performed different MC simulations calculating the total mean-force potential $\Phi(H)$ for increasing values of $\kappa$. Figure \ref{kappa2} shows the effective interaction $\Phi(H)/k_B T$ for  $0.5$ nm$^{-1}$$\leq \kappa \leq 1.25$ nm$^{-1}$.

The observed behavior of the mean-force potential as a function of the ionic strength can be intuitively understood if the potential is explicitly thought of as the difference between the total energy of the two complexes when they are isolated $E(\infty)$ and when they are at a distance $H$ (see Eq. \ref{pairpot1}). For an unscreened potential ($\kappa R_c \ll 1$) each complex interacts with all the charged elements of the opposing complex, hence, the averaged interaction between the two particles results in a function of their overall net charge. On the contrary, when the screening is sufficiently large ($\kappa R_c > 1$), the effective interaction is in practice limited to those parts of the complexes that are sufficiently close each other ($H<\kappa^{-1}<R_c$). The interaction could be hence defined as a \emph{surface-surface} interaction (as opposite to a \emph{particle-particle} one). In these conditions, the local attraction between oppositely charged domains on the two opposing particles can give rise to a net attraction of purely electrostatic nature, even though the particles are, as a whole, like charged.

For the charge ratio employed here ($\xi_s=0.57$), the crossover from repulsive to attractive regime is observed for $2.56<\kappa R_c<3.41$. Above this range a complete destabilization of the dispersion occurs.

At very short distances the effective interaction 'feels' the effect of the steric repulsion due to the overlapping of opposing PE-layers. The soft shell of adsorbed polyelectrolytes enveloping the macroion 'hard sphere', significantly contributes to the interaction only when the two complexes are in close contact.

\begin{figure}[htbp]
\begin{center}
  \includegraphics[width=8.5cm]{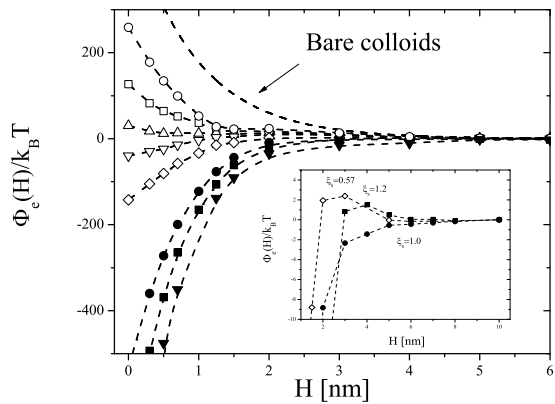}\\
  \caption{The electrostatic component of the meanforce potential $\Phi_e$ between PE-colloid complexes for $\kappa=1$ nm$^{-1}$, $R_c=5$ nm, $Z=-523$ and different values of $\xi_s$. Main panel: Dashed line $\rightarrow$ potential calculated from Eq. \ref{DH2} for bare Z-macroions ($\xi_s=0$); o) $\xi_s=0.11$; $\square$) $\xi_s=0.26$ ; $\triangle$) $\xi_s=0.38$; $\nabla$) $\xi_s=0.46$; $\lozenge$) $\xi_s=0.57$; $\bullet$) $\xi_s=1.0$; $\blacksquare$) $\xi_s=1.2$; $\blacktriangledown$) $\xi_s=1.33$. Inset: potential profiles in the 'intermediate' $H$-region, where an energy barrier between the non neutral complexes arises.}\label{xi_dep}
  \end{center}
\end{figure}

In practice, the effect of the adsorption of the PE-chains is double: the progressive neutralization of the net charge of the complexes that progressively quenches the long range repulsion, and (due to the correlated character of the adsorption and to the presence of a screening) the appearance of a purely electrostatic short range attraction. In Figure \ref{xi_dep} we note the progressive reduction of the repulsion in the electrostatic component of the effective interaction $\Phi_e(H)/k_B T$ as the polyelectrolyte content is increased and, as expected, the rising of a strong short range attraction for $\xi_s \geq 0.46$, well below the isoelectric point ($\xi_s=1$).

In the inset of Figure \ref{xi_dep} three different profiles of the electrostatic interaction $\Phi_e(H)/k_B T$ are shown, to emphasize what occurs below and above the isoelectric point ($\xi_s=1.0$), where there is not monopole-monopole repulsion at all. When a patch attraction is present but the complexes are not neutral, the competition between the residual repulsive interaction and the shorter ranged attraction determines a potential barrier, whose strength depends on the balancing between these two components. The presence of such barrier justifies \cite{Truzzolillo08} the  aggregation behavior observed in several colloid-polyelectrolyte systems regulated by the amount of added polyelectrolyte chains \cite{Bordi04,Bordi04b,Bordi04c,Bordi04d,Bordi05,Bordi05b,Bordi06,Sennato08,Bordi09}.

\begin{figure}[htbp]
\begin{center}
  \includegraphics[width=8.5cm]{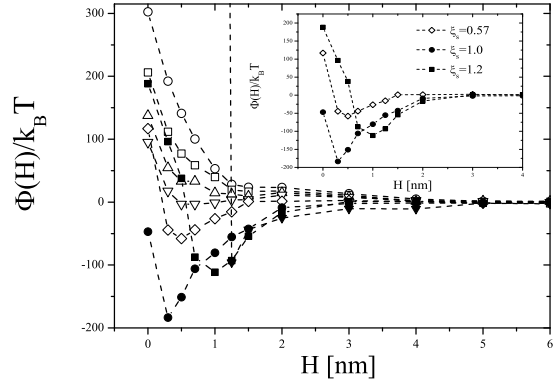}\\
  \caption{Effective interaction between PE-colloid complexes for $\kappa=1$ nm$^{-1}$, $R_c=5$ nm, $Z=-523$ and different values of $\xi_s$. Main panel: o) $\xi_s=0.11$; $\square$) $\xi_s=0.26$ ; $\vartriangle$) $\xi_s=0.38$; $\triangledown$) $\xi_s=0.46$; $\lozenge$) $\xi_s=0.57$; $\bullet$) $\xi_s=1.0$; $\blacksquare$) $\xi_s=1.2$; $\blacktriangledown$) $\xi_s=1.33$. Inset: potential profiles ($\xi_s=0.57, 1.0, 1.2$) in the region of the global minimum.}\label{xi_depB}
  \end{center}
\end{figure}

By adding  a 'steric' contribution to the inter-particle potential, due to the overlapping of PE-layers, we recover the complete effective interaction between the decorated macroions. Figure \ref{xi_depB} shows different profiles of $\Phi(H)/k_B T$ for $0.11\leq\xi_s\leq 1.33$. The global minimum of the potential is the result of the competition between the repulsive steric component and the patch attraction. The position of this minimum is affected by the charge ratio only for high values of $\xi_s$, where the PE-layer cannot be considered flat anymore and the dangling ends protruding from macroion increase the overlap volume of the adsorbed layers on the opposing particles, thus causing a shift of the minimum towards higher values of $H$.

\section{The dipole Moment of the complexes: the Anti-parallel doublets}

As a further investigation of the effects of the interaction between the decorated particles on the conformation and distribution of the adsorbed polyelectrolyte chains, we evaluate the dipole moment of the complexes. It must be noted that for a system whose net charge is different from zero, the dipole moment is not univocally defined but assumes different values depending on the choice of the origin of the coordinates. However, to evaluate a 'degree of asymmetry' of the charge distribution of a  pd-particle we calculate the expression
\begin{equation}\label{dipolecalc}
 \vec{\mu}=\sum_i^{N_q}q_i(\vec{r}_i-\vec{r}_{cc})
\end{equation}
where $N_q$ is the total number of charges $q_i$ on the pd-colloid and $\vec{r}_i$ individuate the position of each charge with respect to a fixed origin (in our case a corner of the box), and
\begin{equation}\label{dipolecalc2}
 \vec{r}_{cc}=\frac{\sum_i^{N_q}|q_i|\vec{r}_i}{\sum_i^{N_q}|q_i|}
\end{equation}
is the baricenter of the charges.

Defining $\vec{d}=\vec{r_{cc+}}-\vec{r_{cc-}}$ as the distance between the center of the positive charges $\vec{r_{cc+}}$ and that of negative charges $\vec{r_{cc-}}$, (which, in our case, coincides with the center of the spherical macroion) Eq. \ref{dipolecalc} can be rewritten, after some algebra, as
\begin{equation}\label{dipolecalc3}
\vec{\mu}=-\frac{2Z\xi_s}{\xi_s+1}\vec{d}
\end{equation}

For an isolated pd-particle, where the charge distribution is, on the average, spherically symmetric, the dipole moment is zero since $\vec{r_{cc+}}=\vec{r_{cc-}}$. However, if the distribution of the adsorbed charges is 'distorted' by an external field, the modulus of $\vec{\mu}$ increases, its value depending on the charge ratio and on the non zero value of $\vec{d}$ (see Eq. \ref{dipolecalc3}).

\begin{figure}[htbp]
\begin{center}
  \includegraphics[width=8.5cm]{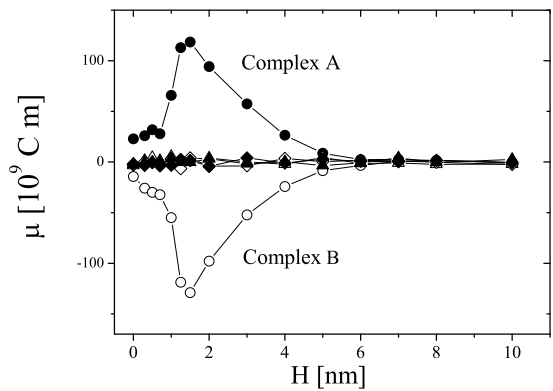}\\
  \caption{The three components of the electrical dipole moment of the complexes A and B shown in Figure \ref{dipoleAB_b} as a function of the distance $H$ between the macroions' surfaces; $\xi_s=0.26$ .}\label{dipolexyz}
  \end{center}
\end{figure}

\begin{figure}[htbp]
\begin{center}
  \includegraphics[width=6.5cm]{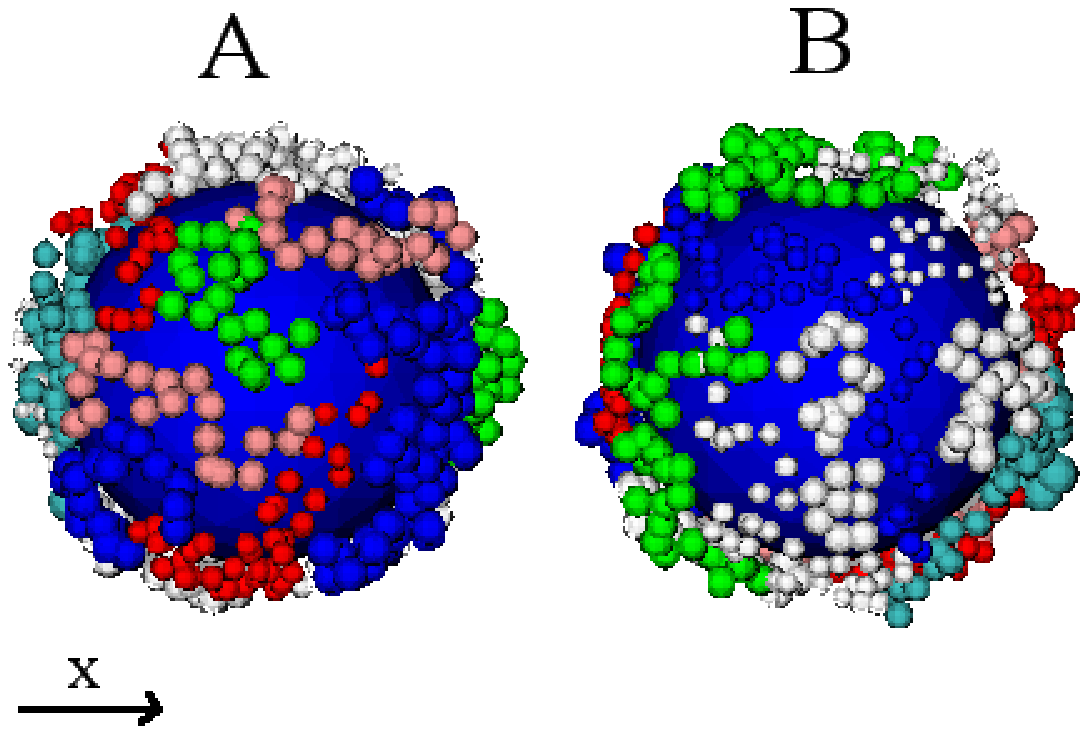}
  \includegraphics[width=8cm]{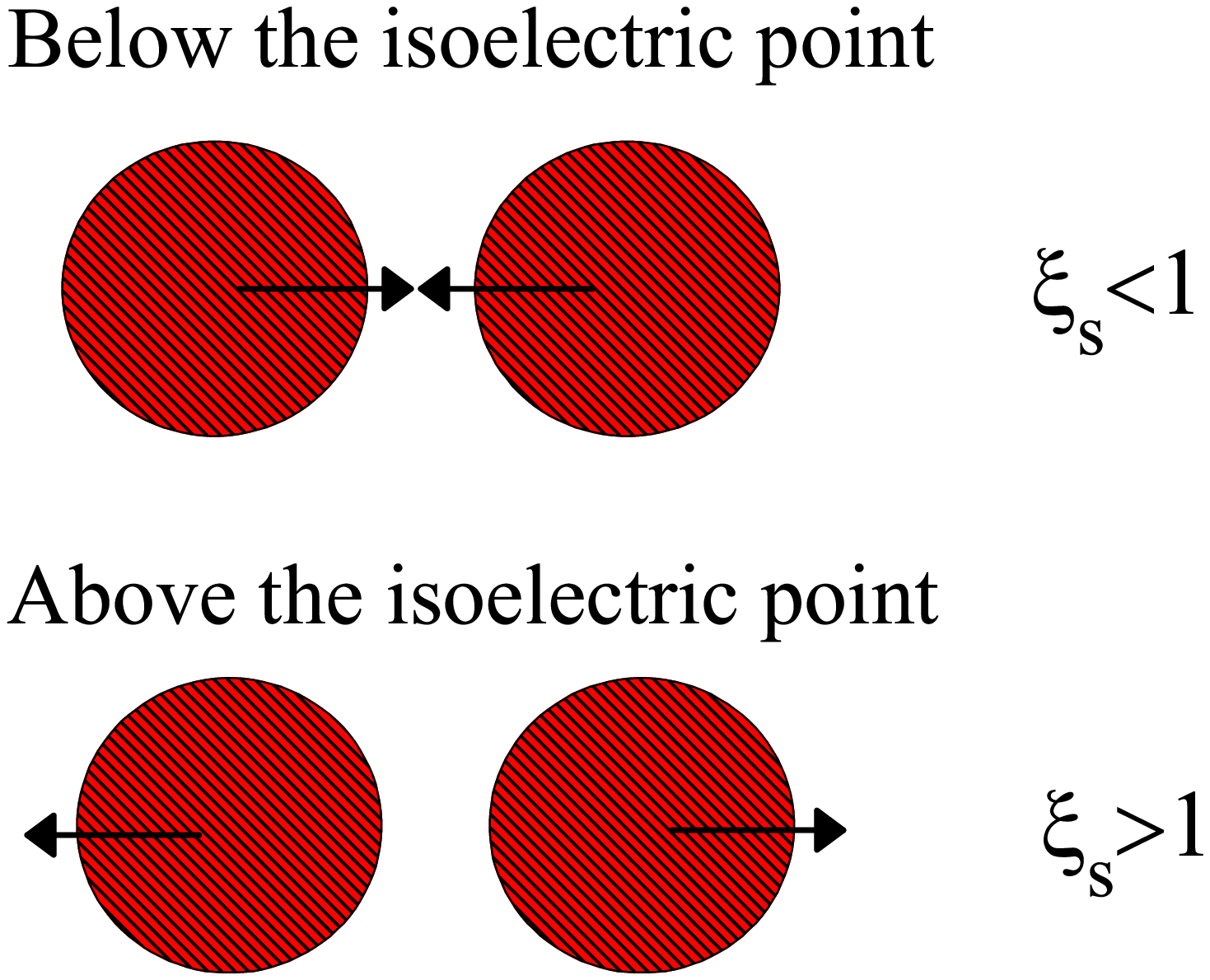}\\
  \caption{
   A snapshot of the labeled complexes (A,B) for $\xi_s=1.0$ and the dipole orientation below and above the isoelectric point.}\label{dipoleAB_b}
  \end{center}
\end{figure}

Due to the symmetry of the system, only the component of the dipole vector along the direction connecting the centers of the two macroions (the x axis with our choice of the coordinates) is affected. For 'intermediate' values of $H$ ($\kappa^{-1}<H<R_c$)  depends on the net charge of the pd-particles. In Figure \ref{dipolexyz} $\mu_x (H)$, $\mu_y(H)$ and $\mu_z(H)$ of two interacting complexes (A and B in Figure \ref{dipoleAB_b}) are shown for $\xi_s=0.26$, $R_c=5$ nm $\kappa=1$ nm$^{-1}$.

The increase of $\mu_x$, for high values of $H$, is due to the asymmetry of the surface charge distribution induced by the opposing particle. At short distances a 'rapid' decrease of $\mu_x$ is observed, deriving from the entropic exclusion of the PE-chains from the contact region, where the steric interaction between monomers and between the monomers and the macroions' surfaces dominates.

\begin{figure}[htbp]
\begin{center}
  \includegraphics[width=8.5cm]{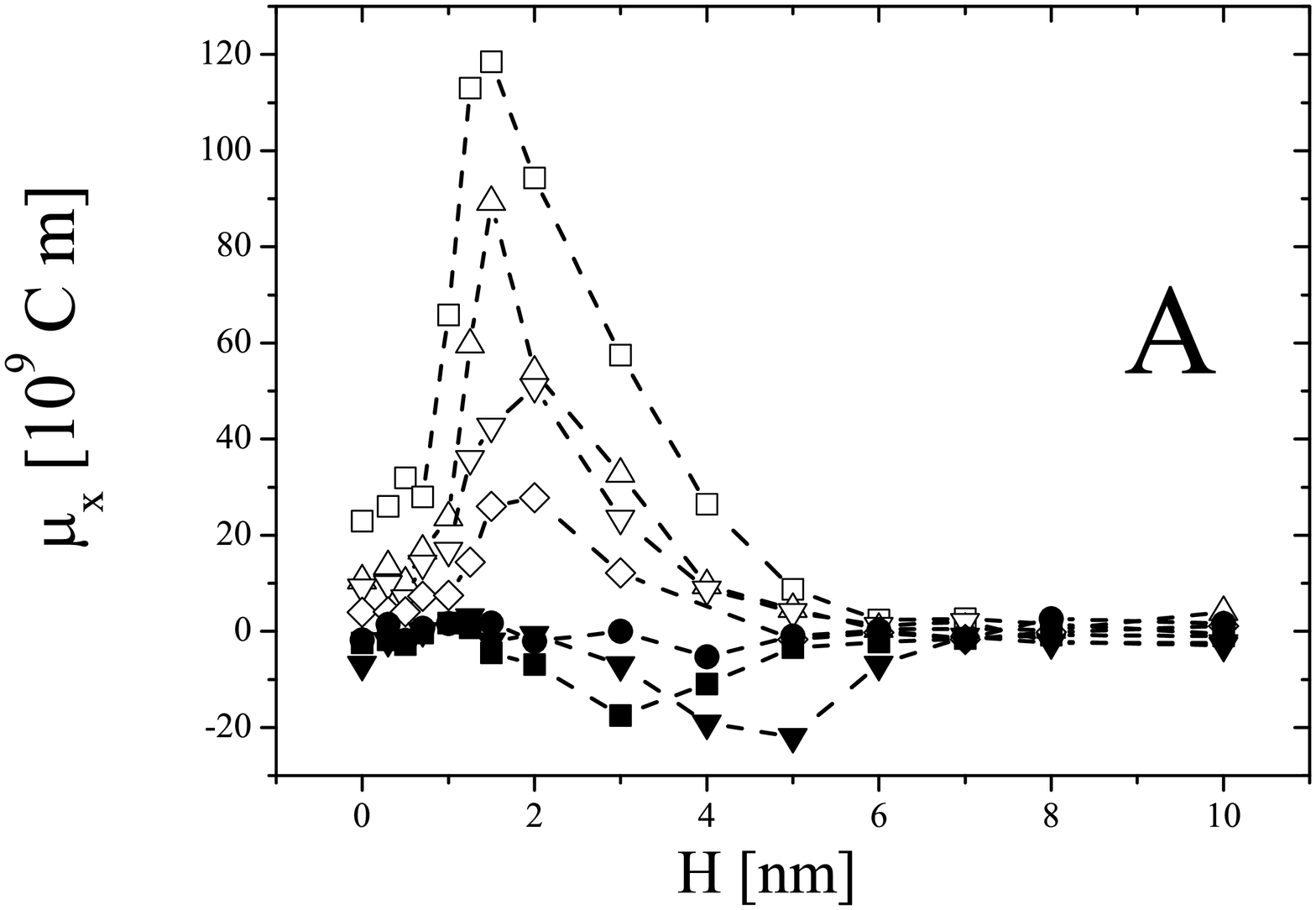}
  \includegraphics[width=8.5cm]{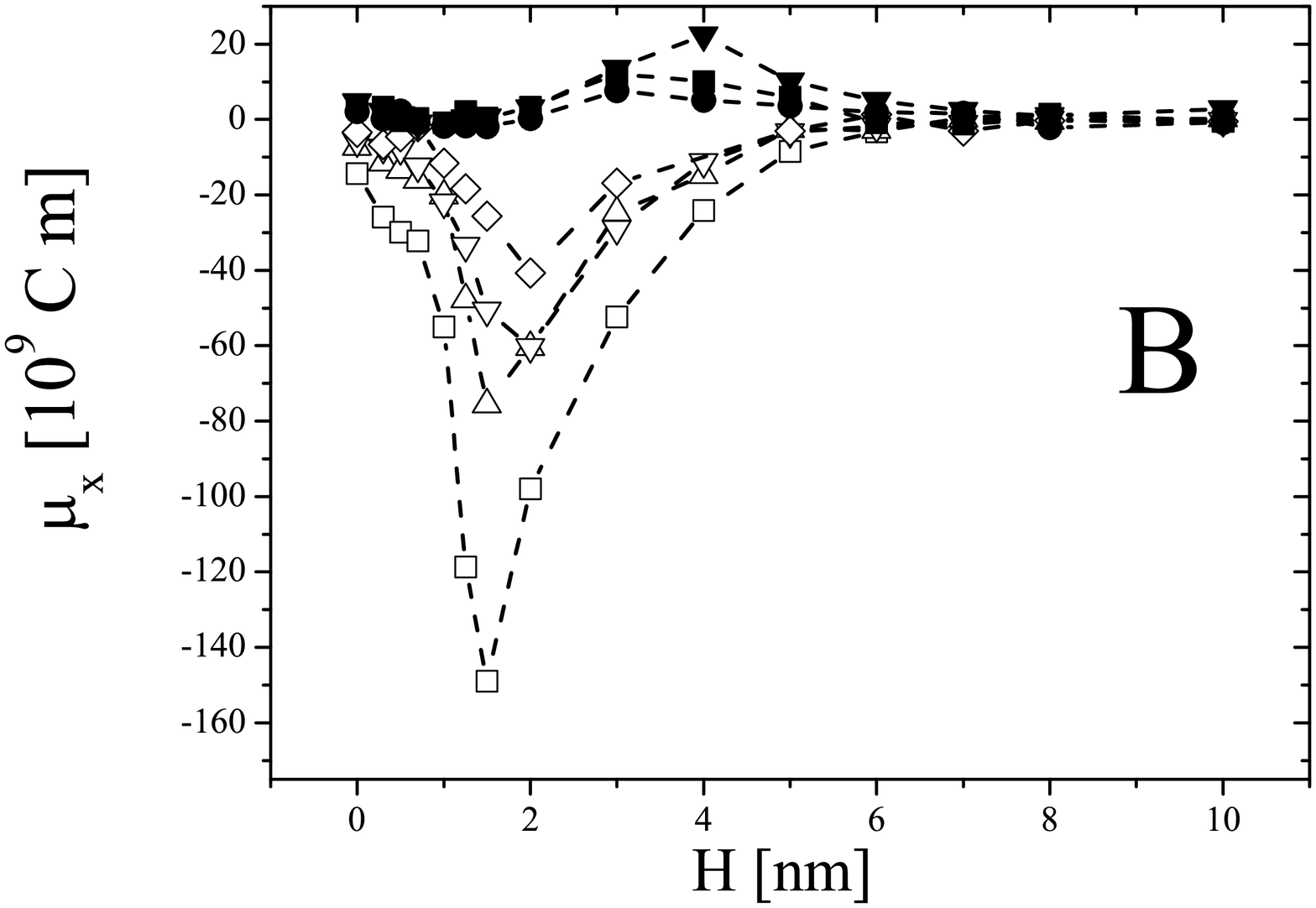}\\
  \caption{A) the profiles of $\mu_x(H)$ for both the complexes (A,B Fig. \ref{dipoleAB_b}): $\Box$) $\xi_s=0.26$ ; $\triangle$ ) $\xi_s=0.38$; $\nabla$) $\xi_s=0.46$; $\lozenge$) $\xi_s=0.57$; $\bullet$) $\xi_s=1.0$; $\blacksquare$) $\xi_s=1.2$; $\blacktriangledown$) $\xi_s=1.33$.}\label{dipoleAB}
  \end{center}
\end{figure}

The effect of the charge ratio  $\xi_s$ on the dipole moment is shown in Figure \ref{dipoleAB}.

An asymmetric distribution of the mobile charges (the adsorbed polyelectrolyte chains) on the macroions can be generated only for distances between the complexes so short that the strength of the interactions is sufficient to promote energy driven diffusion processes that cause a redistribution of the adsorbed chains.
The $x$-components of the dipoles are specular and their signs exchange in correspondence of the isoelectric point. In this way, the eventual aggregation of the two complexes occurs with the \emph{anti-parallel} orientation of the dipoles (+- or -+ depending on wether the aggregation occurs below ($\xi_s<1$) or above ($\xi_s>1$) the neutralization condition). This anti-parallel orientation of the dipole doublet signals the absence of a long-range dipole-dipole attractive contribution to the mean force inter-particle potential.
Such findings give further support to the hypothesis that the attractive force that drives the aggregation of like charged pd-colloidal particles arises from the charge mismatch between polyelectrolyte and polyelectrolyte-free domains on the opposing macroions' surfaces.

\section{Final remarks}
We studied the effective interaction and the induced charge asymmetry of PE-colloid complexes within the Debye-H\"{u}ckel approximation for the case where the monomer size is small compared with that of the adsorbing macroion and the surface charge density of bare colloids is characteristic of the extensively investigated spherical DOTAP liposomes
\cite{Bordi04,Bordi04b,Bordi04c,Bordi04d,Bordi05,Bordi05b,Bordi06,Bordi09}.
The interaction between non-neutral PE-decorated particles is purely repulsive for large screening length $\kappa R_c \ll 1$ and it is well-described by a Debye-H\"{u}ckel potential. The interaction is sensibly modified as the screening is increased until a short ranged attractive component appears.

From the competition between this short range attraction and the long range electrostatic repulsion due to the residual net charge of the like-charged complexes  a global minimum in the pair potential arises whose position depends on the  thickness of the adsorbed PE-layers. By adding chains on the macroions' surface the short range electrostatic component of the potential switches from repulsive to attractive well-below the neutralization condition ($\xi_s\simeq 0.46$). For increasing values of the charge ratio $\xi_s$, the adsorbed chains concomitantly reduces their typical dimension and form thick and spatially non homogeneous layers.
Interestingly, the dipole moments of two approaching complexes are always anti-parallel (except at the neutralization point where all the dipoles' components are zero)  so that the condensation of these complexes can not be attributed to dipole-dipole attractive interactions.

\newpage

\newpage
\end{document}